\documentclass[%
amsmath,amssymb,
aps,
prb,
twocolumn
]{revtex4-2}

\usepackage{graphicx}
\usepackage{dcolumn}
\usepackage{bm}
\usepackage{xcolor}
\usepackage[caption=false]{subfig}
\usepackage{tikz}
\usepackage{xfrac}
\usepackage{blindtext}
\usepackage{times}

\renewcommand{\vec}[1]{{\boldsymbol #1}}

\definecolor{darkblue}{HTML}{004D6B}
\definecolor{darkred}{HTML}{8c1515}

\usepackage{hyperref}
\hypersetup{
	colorlinks=true,
	urlcolor=darkred,
	citecolor=darkblue,
	linkcolor=darkred,
	breaklinks
}

\usepackage{physics}
\usepackage{bbold} 

\begin{document}
	
	\title{Thermalization and hydrodynamic long-time tails in a Floquet system}
	\author{Anne Matthies$^1$}
	\author{Nicolas Dannenfeld$^1$}
	\author{Silvia Pappalardi$^1$}
	\author{Achim Rosch$^1$}
	\affiliation{$^1$Institute for Theoretical Physics, University of Cologne, 50937 Cologne, Germany}
	
	\date{\today}
	
	\begin{abstract}
We systematically investigate whether classical hydrodynamic field theories can predict the long-time dynamics of many-particle quantum systems. 
We study both numerically and analytically the time evolution of a chain of spins (or qubits) subjected to stroboscopic dynamics. The time evolution is implemented by a sequence of local and nearest-neighbor gates that conserve the total magnetization. The long-time dynamics of such a system is believed to be describable by a hydrodynamic field theory, which, importantly, includes the effect of noise.
Based on a field theoretical analysis and symmetry arguments, we map each operator in the spin model to the corresponding fields in hydrodynamics. This allows us to predict which expectation values decay exponentially and which decay with a hydrodynamic long-time tail. We illustrate these findings by studying the time evolution of all 255 Hermitian operators that can be defined on four neighboring sites. All operators not protected by hydrodynamics decay exponentially, while the others show a slow hydrodynamic decay. While most hydrodynamic power laws seem to follow the analytical predictions, we also discuss cases where there is an apparent discrepancy between analytics and the finite-size numerical data.
	\end{abstract}
	
	\maketitle

	Understanding how classical and quantum systems of many particles reach thermal equilibrium has been a venerable question for many decades \cite{Polkovnikov2011,Eisert2015,Gogolin2016}. 
The universal long-time and large-distance dynamics of large classes of both classical and quantum systems are well understood via \emph{hydrodynamics}, which describes the emergent behavior of conserved densities protected by continuous symmetries \cite{Lebowitz1988, Forster1994}. 
Such theories are believed to apply to thermalizing systems, which approach a finite-temperature thermal state when left alone~\cite{Mori2018}. 
There are several successful approaches to hydrodynamics, ranging from fluctuating hydrodynamics \cite{Spohn1991, martin1973statistical, forster1977large,spohnChain} and macroscopic fluctuation theory \cite{Bertini2015} to the effective field theory of diffusion \cite{liu2018lectures,jensen2018panoply, haehl2018effective, crossley2017effective}. These frameworks possess remarkable predictive power in describing the long-time behavior of generic hydrodynamic systems, including corrections at late times \cite{Michailidis2024} or nonlinear effects \cite{delacretaz2024nonlinear}. 

Hydrodynamic field theories are highly universal, as they depend only on the conservation laws in the system \cite{Forster1994}. In its simplest setting, there is a single diffusive conservation law. In time-translationally invariant Hamiltonian systems, the energy is always conserved. In a generic spin system with spin-orbit interactions, for example, there is no other conservation law besides energy, and one obtains only a single diffusive mode. Two coupled diffusive conservation laws arise, for example, in metals, where, besides energy, charge is also conserved. In fluids, there is additionally momentum conservation, and the Navier-Stokes equations (with added noise terms) define the hydrodynamic field theory. The Navier-Stokes equations in one and two dimensions
differ from simpler hydrodynamic theories because non-linearities, arising from the coupling of momentum currents to charge or energy, are relevant in the renormalization group sense and thus can dominate the behavior at long time and length scales. This leads to dynamics in the Kardar–Parisi–Zhang (KPZ) universality class \cite{spohnChain,Popkov_2016,Spohn2020a}. The framework of hydrodynamics can also be extended to describe Goldstone modes in systems with spontaneously broken continuous symmetries  \cite{Forster1994} and even to 
 (classical or quantum) integrable many-particle systems, leading to the so-called generalized hydrodynamics \cite{bertini2016transport,castroalvaredo2016emergent,bulchandani2017solvable,Huebner2024}.

While hydrodynamics is a purely classical theory, it can be used directly as an effective description of many-body quantum systems at large scales \cite{mukerjee2006statistical, Lux2014, bertini2021finite}. 
A prototypical example is the so-called quantum quench, where the unitary evolution of local observables from an initial pure state exhibits slow power-law relaxation, $\sim t^{-\alpha}$, even in translationally invariant quantum systems  \cite{Lux2014}. These  `hydrodynamic long-time tails'  arise as it takes a very long time to change the amplitude of fluctuations of conserved densities.  
Furthermore, hydrodynamics enters in a variety of different quantum phenomena in chaotic systems, ranging from entanglement entropy dynamics \cite{Rakovszky2019,huang2020dynamics, Zhou2020diffusive, vznidarivc2020entanglement, rakovszky2021entanglement}, the scrambling of quantum operators \cite{bohrdt2017scrambling, khemani2018operator,rakovszky2018diffusive, cheng2021scrambling}, spectral statistics \cite{Friedman2019spectral, Roy2020, Roy2022,Kumar2024}  and the eigenstate-thermalization hypothesis \cite{Capizzi2024}.  A recent study investigated how non-linear hydrodynamics can be captured in the framework of the eigenstate thermalization hypothesis, which allowed to predictions of the late-time behavior of time-ordered free cumulants in the thermodynamic limit \cite{wang2025}. 

 There is only a limited number of numerical studies~\cite{Ye2020,rakovszky2018diffusive,bohrdt2017scrambling} of hydrodynamic long-time tails after a quantum quench starting from Ref.~\cite{Lux2014}. A recent extensive numerical study by Maceira and Läuchli~ \cite{Maceira2024} of quenches in an energy-conserving  Hamiltonian system with up to 34 spins studied the relaxation at very long times and, surprisingly, found an exponential relaxation of local observables with a relaxation time growing approximately linearly in system size.

The emergence of hydrodynamics in quantum systems is interesting in its own right, but it has regained significant attention recently due to experimental progress in the field of quantum simulation. Remarkable experiments on cold atom platforms have demonstrated the power of analog quantum simulators \cite{Trotzky2012, Kaufman2016, Erne2018, Zhou2022, Schneider2012}, which allow for experimental exploration of the continuous time evolution of a given Hamiltonian \cite{Zu2021, Wienand2024}.
More recently, with the advent of quantum computing, significant advances have been made in \emph{digital simulations of time evolution}, where discrete dynamics are implemented via unitary gates acting locally or on nearest neighbors \cite{Georgescu2014quantum, Yang2023}. In this case, even in the absence of energy conservation, hydrodynamics can still emerge if another conservation law is present, such as the conservation of total magnetization. 
This case has been studied, for instance, for local random quantum circuits, 
 \cite{Rakovszky2019,huang2020dynamics, Zhou2020diffusive, vznidarivc2020entanglement, rakovszky2021entanglement, khemani2018operator,rakovszky2018diffusive, Friedman2019spectral, vznidarivc2020entanglement, Richter2023transport, Jonay2024slow, turkeshi2024quantum}, where the presence of randomness allowed for exact results.  Naturally, these developments raise the question of whether, at large times, i.e., for large circuit depth, the discrete-time evolution in quantum circuits is encoded in hydrodynamic descriptions. 

In this work, we perform a detailed analysis of how hydrodynamics emerges in a clean Floquet system with a discrete translational symmetry. Specifically,  we demonstrate how fluctuating hydrodynamics can be used to predict the long-time tails that arise in the quenched dynamics of generic observables $ A$
\begin{equation}
	\label{eq_At}
	\langle\psi(t_n) | A |\psi(t_n)\rangle\ ,
\end{equation}
where $\ket{\psi(t_n)} = U_{\rm FL}^n\ket{\psi_0}$ is the state evolved at discrete times $t_n$ with a Floquet operator $U_{\rm FL}$ which conserves the total magnetization. A combination of two effects dictates these predictions: 1) the interplay of the hydrodynamic modes with symmetries and 2) the role of the initial states. While previous studies on hydrodynamics have often focused on correlation functions of conserved densities, we aim to demonstrate that hydrodynamic predictions apply to a much broader set of observables.
Thus, we study the time evolution of all 255 possible Hermitian operators $A$ which can be defined as products of $X_i$, $Y_i$, $Z_i$, and the identity matrix on four different sites. For each of these operators, we use field theory and symmetry arguments to predict their long-time dynamics. While 128 of them are not protected by any symmetry and hence decay exponentially, we provide a comprehensive analysis of how to understand the impact of the hydrodynamic modes from a derivative expansion.
We also discuss the role of the initial states in determining the leading behavior at long times. 

Finally, we numerically confirm the validity of our predictions by studying the exact time evolution of time-dependent observables, as described in Eq.\eqref{eq_At}, for systems with up to $N=28$ qubits. We also discuss the influence of finite-size effects, which are mostly exponentially suppressed with  $2^{-N/2}$. These effects may hide some of the long-time tails in cases where they possess a numerically small prefactor.

Our results highlight the predictive power of hydrodynamics in describing the long-time behavior of small strings of observables and demonstrate the importance of symmetries in characterizing long-time dynamics in digital quantum time evolution.

The rest of the paper is organized as follows. In Sec.~\ref{sec:model}, we introduce the model and discuss the different initial states under analysis. In Sec.~\ref{sec:hydro}, we provide a summary of fluctuating hydrodynamics and classify the various operators of the spin model to predict their long-time tails. Finally, in Sec.~\ref{sec:num} we present the numerical results. We conclude in Sec.~\ref{sec:conclusion} with a summary and a discussion of the possible failures of hydrodynamics.

\section{Model and Conventions}\label{sec:model}
We are interested in a generic thermalizing quantum spin system with only one conserved quantity. 
Our goal is to use the simplest possible setting for equilibration, which can also be efficiently simulated numerically. Thus, we chose a system without energy conservation, which equilibrates to an infinite temperature state.
To break energy conservation, we consider a stroboscopically driven system, which has the advantage that it can be simulated numerically more easily (without possible Trotter errors). The time evolution is generated by repeatedly applying a Floquet operator
 $U_{\text{FL}}$ 
\begin{equation} 
	|\psi_n \rangle = \left( U_{\text{FL}}\right)^n |\psi_0\rangle.\label{eq:timeevolution}
\end{equation}
 where $U_{\text{FL}}$ is given by
 \begin{equation}\label{eq:U}
	U_{\text{FL}}=e^{-i H_\gamma}e^{-i H^\text{even}_\beta }  e^{-i H^\text{odd}_\beta }e^{-i H_\alpha}.
 \end{equation}

$H_\alpha$ and $H_\beta$ describe a stroboscopic version of an integrable spin-$1/2$ $XXZ$-chain
	\begin{align}
H_{\alpha} =&-\alpha\sum_i   Z_i Z_{i+1} \nonumber \\
H^\text{odd}_\beta=&-\beta \sum_{i} \left(X_{2 i-1} X_{2 i}+Y_{2 i-1} Y_{2 i}\right)\nonumber \\ 
H^\text{even}_\beta=&-\beta \sum_{i} \left(X_{2 i} X_{2 i+1}+Y_{2i} Y_{2 i+1}\right),
	\end{align}
where $X_i$, $Y_i$, $Z_i$ describe the Pauli matrices at site $i$ and we consider periodic boundary conditions.  For the $XX$ and $YY$ couplings, even and odd links are treated separately as they do not commute with each other. This allows for an implementation using only 2-qubit gates.
 The last term ($H_\gamma$) breaks integrability
	\begin{align}
	H_\gamma=&-\gamma \sum_i Z_i Z_{i+2}.
	\end{align}
This model only conserves the total $z$-magnetization ${M_z=\sum_i Z_i}$, i.e.
\begin{align}
[M_z,U_\text{FL}]=\left[\sum_i Z_i,U_\text{FL}\right]=0.\label{eq:conserved}
\end{align}
Throughout the paper, we chose the parameters as $\alpha=2$, $\beta=0.25$, and $\gamma=1$.
Experimentally, such time evolutions can be realized directly on quantum computers, where, however, errors and noise terms provide extra complications. As our main goal is to validate and test hydrodynamic approaches, we ignore such effects in the following.

We consider three different initial states 
\begin{align}
  |\psi_0^{(1)}\rangle&=\left(|\uparrow\rangle_Z | \downarrow\rangle_Z\right)^{\otimes N/2} \nonumber \\
 |\psi_0^{(2)}\rangle&= |+\rangle_X^{\otimes N} \nonumber \\
 |\psi_0^{(3)}\rangle&= \left(|+\rangle_X|+\rangle_Y|-\rangle_X|-\rangle_Y \right)^{\otimes N/4},
\label{eq:initial}
\end{align} 
describing an initial state $ |\psi_0^{(1)}\rangle$ with staggered magnetization in the z-direction,  
a state  $ |\psi_0^{(2)}\rangle$ with a magnetization in $+$ x-direction, and a state $ |\psi_0^{(3)}\rangle$
where the magnetization rotates in the xy-plane. The third state has been chosen as it implements a large and finite spin current in its initial state, $\langle \psi_0^{(3)}| X_i Y_{i+1}-Y_i X_{i+1} |\psi_0^{(3)}\rangle=1$ independent of $i$.  Importantly, all three states are characterized by a discrete translational invariance.

Our goal is to compute and predict the time-dependence of expectation values of observables $A$,  for long times and large system sizes based on the time evolution of Eq.~\eqref{eq:timeevolution}. We use the notation
\begin{align}
\langle A(t_n) \rangle^{(m)}\equiv \langle \psi_0^{(m)}| (U_\text{FL}^\dagger)^n \, A\,\, (U_\text{FL})^n |\psi_0^{(m)}\rangle
\end{align}
to denote expectation values of operators computed using  $|\psi_0^{(m)}\rangle$ as an initial state.

\section{Fluctuating Hydrodynamics}\label{sec:hydro}
\subsection{Fluctuating hydrodynamics as a field theory}
Hydrodynamics is a classical effective field theory believed to capture the long-distance, large-time dynamics of interacting quantum or classical many-body systems close to equilibrium~\cite{Lebowitz1988,Forster1994}. 
It builds on the assumption that generic many-particle systems (with the notable exception of integrable systems, many-body localized systems, and various types of glasses) relax towards equilibrium starting from arbitrary excited states. For almost all degrees of freedom, this relaxation is expected to be fast and exponential, and the only exceptions are modes protected by continuous symmetries (including Goldstone modes if a continuous symmetry is spontaneously broken). We denote the time scale for local equilibration of these modes by $\tau^\text{loc}$. In the Floquet system studied by us, there is only a single continuous symmetry, generated by the total magnetization operator $M_z=\sum_i Z_i$, see Eq.~\eqref{eq:conserved}. Within the hydrodynamic field theory, we identify $M_z$ with $\int d x\, m(x,t)$, where $m(x,t)$ is the local magnetization density.
As magnetization cannot be generated or destroyed locally, $m(x,t)$ is a slow mode. As there is no other conservation law, it defines the only slow mode in our system.
Our study can be viewed as a test of the assumption underlying hydrodynamics.

The hydrodynamic equation is obtained from the continuity equation for $m(x,t)$ 
\begin{align}
\partial_t m + \nabla j &= 0
\end{align}
and an expansion of the current $j$ in terms of $m(x,t)$
\begin{align}
j &\approx -D_0 \nabla m - \eta + D_2 m^2 \nabla m + D_3 \nabla^3 m +D_4 \nabla \partial_t m \dots \label{eq:current}
\end{align}
where $D_0$ is the (magnetic) diffusion constant, while $D_2, D_3, \dots$ describe non-linear and non-local (in space and time) corrections to the diffusion law. Here, we used the fact that $j$ is odd under both inversion and the transformation $m \to -m$. Importantly, fluctuating hydrodynamics also includes a random noise term, $\eta(x,t)$ with $\langle \eta \rangle=0$, arising physically from other, non-hydrodynamic degrees of freedom. 
Using a Gaussian noise field $\xi$ with $\langle \xi(x,t) \xi(x', t') \rangle =\delta(x-x') \delta(t-t')$,
We can use a Taylor expansion similar to Eq.~\eqref{eq:current} to express
\begin{align}
\eta(x,t) \approx & \sqrt{2 C_0} \xi(x,t) \\
& +  C_2 m^2 \xi+C_3 \nabla^2 \xi 
+ C_4  (\xi^3-3 \langle \xi^2  \rangle \xi) + \dots \nonumber
\end{align}
where the second line includes symmetry-allowed corrections for non-linear, non-local, and non-Gaussian noise. 
The linearized noisy diffusion equation to leading order in the gradient expansion becomes
\begin{align}\label{eq:linear}
&\partial_t m = D_0 \nabla^2 m + \nabla \eta \\
& \langle \eta(x,t) \eta(x', t') \rangle \approx 2 C_0 \delta(x-x') \delta(t-t') \nonumber
\end{align}
with $\langle \eta(x,t) \eta(x', t') \rangle \approx 2 C_0 \delta(x-x') \delta(t-t')$ . The noise amplitude is (by fluctuation dissipation theorems) proportional to the diffusion constants and the magnetic susceptibility. 

The rescaling $x \to \tilde x = x/\lambda$, $t \to \tilde t = t/\lambda^2$, 
$\eta \to \tilde \eta = \eta \lambda^{(d+2)/2}$, $m\to \tilde m=m \lambda^{d/2}$ leaves the linearized theory in $d$ spatial dimension invariant.
According to the scaling analysis, all corrections to linear hydrodynamics are irrelevant in the renormalization group sense as they are suppressed for large $\lambda$ when using the rescaled coordinates. More specifically, in $d=1$ we obtain
\begin{align}
D_2 \sim C_2 \sim \frac{1}{\lambda},\quad
 D_3\sim D_4 \sim C_3 \sim \frac{1}{\lambda^2}, \quad C_4 \sim \frac{1}{\lambda^3}\label{eq:scalingNonLin}
\end{align}
Thus, their effect vanishes in the long-time limit (see below), and we will need them only to compute subleading corrections to the long-time dynamics of observables.

After rescaling, observables can be expressed in terms of the rescaled variables. As we are interested in the value of observables measured at time $t$ in the limit of long times, it is useful to choose $\lambda=\sqrt{t}$ such that $\tilde t=1$ and then expressing everything in the rescaled variables. For  $d=1$, this leads to the following rules for power-counting used heavily in the following
\begin{align}\label{eq:powercounting}
\nabla_x &= \frac{\nabla_{\tilde x}}{t^{1/2}}\sim \frac{1}{t^{1/2}}, \quad \partial_t =\frac{\partial_{\tilde t}}{t}\sim \frac{1}{t}, \nonumber \\
 m &=\frac{\tilde m}{t^{1/4}}\sim \frac{1}{t^{1/4}}, \quad  \eta = \frac{\tilde \eta}{t^{3/4}}\sim \frac{1}{t^{3/4}}.
\end{align}

Such scaling  arguments can be used to show that the effects arising from the leading non-linear 
 terms $D_2$ and $C_2$ will be suppressed by powers of $1/\sqrt{t}$ relative to leading terms, see Eq.~\eqref{eq:scalingNonLin}.

In the following, we show in a simple calculation how long-time tails emerge in the linearized theory.
The magnetization at time $t$ is obtained from a straightforward solution of the linearized diffusion equation, Eq.~\eqref{eq:linear}
\begin{widetext}
	\begin{align}
		m(x,t)=\int_{-\infty}^\infty dx' G(x-x',t)m(x',0)+\int_{-\infty}^\infty dx' \int_{0}^{t} dt'G(x-x',t-t')\nabla \eta(x',t')
	\end{align}
	where $G(x,t)=\theta(t) e^{-x^2/(4 D_0 t)}/(4 \pi D_0 t)^{1/2}$ is the Green's function of the diffusion equation with Fourier transformation $G_k(t)=\theta(t) e^{-D_0 k^2 t}$.
We describe the noise by  $\langle \eta(x,t)\eta(x',t')\rangle = \delta(t-t') C^\eta(x-x')$. Furthermore, we assume that the initial correlations at $t=0$ are given by $\langle m(x_1,0)m(x_2,0) \rangle =C(x_1-x_2)$, where $C$ is a function that depends on the initial state. This allows us to calculate the equal-time correlation function  at time $t$
	\begin{align}
		\langle m(x,t)m(\tilde{x},t)\rangle &=\int_{-\infty}^\infty dx_1\int_{-\infty}^\infty dx_2 G(x-x_1,t)G(\tilde{x}-x_2,t)\langle m(x_1,0)m(x_2,0) \rangle\nonumber \\
		& +\int_{-\infty}^\infty dx_1\int_{-\infty}^\infty dx_2 \int_{0}^{t} dt_1\int_{0}^{t} dt_2 G(x-x_1,t-t_1)G(\tilde{x}-x_2,t-t_2)\langle \nabla \eta(x_1,t_1)\nabla \eta(x_2,t_2) \rangle\nonumber \\
	& = \int \frac{dk}{2\pi} G_k(t) G_{-k}(t) C_k e^{i(x-\tilde{x})k}+ \int_0^t dt_1 \int \frac{dk}{2\pi} k^2 G_k(t-t_1) G_{-k}(t-t_1) C^\eta_k e^{i(x-\tilde{x})k}\nonumber \\
	& = \int \frac{dk}{2\pi} e^{-2 D_0 k^2 t} C_k e^{i(x-\tilde{x})k}+ \int \frac{dk}{4\pi D_0} 
	\left(1- e^{-2 D_0 k^2 t} \right) C^\eta_k e^{i(x-\tilde{x})k}
\nonumber \\
	&=\frac{C^\eta(x-\tilde x)}{2 D_0} +\theta(t) \frac{e^{-(x-\tilde{x})^2/(8D_0t)}}{\sqrt{8 \pi D_0 t}} \left[ \left(C_{k=0}-\frac{C^\eta_{k=0}}{2 D_0}\right) + \frac{(4 D_0 t - (x-\tilde{x})^2)}{32 D_0^2 t^2} \left(C''_{k=0}-\frac{C''^\eta_{k=0}}{2 D_0}\right)+\dots \right]\label{eq:correlationDecay1}
	\end{align}
	\end{widetext}
where we used that $\langle m(x) m(x')\rangle=\langle m(x') m(x)\rangle$ and therefore $C(x)=C(-x)$ for a translationally invariant system. Thus, the Fourier transform $C_k=C_{-k}$ has no odd components in $k$ independent of the symmetries of the initial state.
Here, $\frac{C^\eta(x-\tilde x)}{2 D_0}$ can be identified with the steady-state correlation function, which is obtained for $t\to \infty$.
Hydrodynamic long-time tails arise whenever the initial correlation functions and the steady-state correlation function are different. If the $k=0$ correlations, $C_{k=0}=\int C(x) dx$, of the initial state differ from the correlations $\frac{C^\eta_{k=0}}{2 D_0}$,
one obtains, for example, 
\begin{align}
	\label{eq:m2}
\langle m(x,t)^2\rangle \sim \frac{1}{\sqrt{t}},\qquad 
\langle (\nabla m(x,t))^2\rangle \sim  \frac{1}{t^{3/2}}.
\end{align}
Only by diffusive processes the fluctuations of the magnetization can be readjusted, which leads to pronounced long-time tails.

\subsection{Mapping  to hydrodynamics}
To use conventional units for space and time, we first introduce a lattice constant $a$ and also associate each unitary with a time scale $T$, which is its duration. We denote the position of spin $j$ by $x_j=a j$ and the time after the application of $n$ unitaries by $t_n=n T$. As $T$ and $a$ are simply conventions, their value is irrelevant (and can also be set to $1$). We employ a Heisenberg picture for operators $A$ with 
$A(t_n)=(U^\dagger)^n A U^n$.

A mapping of the quantum system to fluctuating hydrodynamics is only possible in the limit $t_n \gg \tau^\text{loc}$, where $\tau^\text{loc}$ is the local equilibration time, which gives an upper limit of the decay-time of all non-hydrodynamic modes. 
The existence of such a timescale is an assumption, which we will test numerically.

The first target of the hydrodynamic equation is to compute expectation values of products of operators $Z_i$. We will focus on equal-time expectation values after initializing the system with a wave function $|\Psi\rangle$ with $
\langle A(t_n) \rangle = \langle \psi| A(t_n) |\psi\rangle$. 
Thus, hydrodynamics should predict in the long-time limit
\begin{align}
\langle Z_{i_1}(t_n) & Z_{i_2}(t_n) \dots Z_{i_m}(t_n)  \rangle
\\ & = 
\langle m(x_{i_1},t_n)  m(x_{i_2},t_n) \dots  m(x_{i_m},t_n) \rangle \nonumber
\end{align}
where $\langle \dots \rangle$ on the left-hand side denotes the quantum mechanical expectation value $\langle \psi_0| \dots |\psi_0\rangle$. In contrast, $\langle \dots \rangle$ on the right side is computed from a solution of the nonlinear hydrodynamic equation using an average over all noise configurations. Here, the initial conditions should be set by matching the correlations and their time derivatives at an initial time $t_\text{ini} \gg \tau^\text{loc}$.

While the mapping of products of $Z_i$ to the slow mode $m(x)$ follows from the definition of $m$, for any other operator, such a mapping is not straightforward. However, we can expect that for an operator $A_i(t_n)$, where $i$ denotes the position, a Taylor expansion exists of the following form 
\begin{widetext}
\begin{align}\label{eq:TaylorOp}
\langle A_i(t_n) \rangle \approx& \Bigl \langle \gamma^{(0,0)}+\int \! dx_1 dt_1 \,\gamma^{(1,0)}_{x_1-x_i,t_1-t_n} m(x_1,t_1) +\int  \! dx_1 dt_1 \,\gamma^{(0,1)}_{x_1-x_i,t_1-t_n} \eta(x_1,t_1)\nonumber\\
&+\int dx_1 dt_1 dx_2 dt_2\, \gamma^{(1,1)}_{x_1-x_i,x_2-x_i,t_1-t_n,t_2-t_n} m(x_1,t_1) \eta(x_2,t_2)\nonumber\\
&+ \int  dx_1 dt_1 dx_2 dt_2\,  \gamma^{(2,0)}_{x_1-x_i,x_2-x_i,t_1-t_n,t_2-t_n} m(x_1,t_1) m(x_2,t_2)\nonumber\\
&+ \int dx_1 dt_1 dx_2 dt_2\,    \gamma^{(0,2)}_{x_1-x_i,x_2-x_i,t_1-t_n,t_2-t_n} \eta(x_1,t_1) \eta(x_2,t_2) +\dots \Bigr \rangle 
\end{align}
where the function $\gamma^{(i,j)}$ is expected to decay exponentially as a function of distance and time. 

Note that the coefficients of the Taylor expansion are independent of the initial state of the system (which is encoded in the initial condition of the hydrodynamic equation as discussed above).

As a next step, one can then perform a Taylor expansion of $m(x_j,t_j)$ around the point $x_i, t_n$ to arrive at
\begin{align}
\langle A_i(t_n) \rangle  \approx & \Bigl \langle \gamma^{(0,0)}+\gamma^{(1,0)}_0 m(x_i,t_n)+\gamma^{(1,0)}_1 \nabla m(x_i,t_n) +\gamma^{(1,0)}_2 \partial_t m(x_i,t_n) +\dots +\gamma^{(0,1)}_0 \eta(x_i,t_n) +\gamma^{(0,1)}_1 \nabla \eta(x_i,t_n) + \dots \nonumber \\
&+\gamma^{(1,1)}_0 m(x_i,t_n)\eta(x_i,t_n)  +\gamma^{(1,1)}_1 \nabla (m(x_i,t_n)\eta(x_i,t_n)) + \dots +\gamma^{(2,0)}_0 (m(x_i,t_n))^2 +\gamma^{(2,0)}_1 \nabla (m(x_i,t_n))^2+\dots \nonumber \\
&+\gamma^{(0,2)}_0 (\eta(x_i,t_n))^2 +\gamma^{(0,2)}_1 \nabla (\eta(x_i,t_n))^2+\dots \Bigr \rangle \nonumber \\
=&\sum_{i,j} \Bigl  \langle \gamma^{(i,j)}_0 m(x_i,t_n)^i \eta(x_i,t_n)^j +\gamma^{(i,j)}_1 \nabla (m(x_i,t_n)^i \eta(x_i,t_n)^j)+\gamma^{(i,j)}_2\partial_t (m(x_i,t_n)^i \eta(x_i,t_n)^j)+\dots \Bigr \rangle,  \label{eq:taylor}
\end{align}
\end{widetext}
where the real-valued prefactors can be computed from the functions $\gamma^{(i,j)}$ defined in Eq.~\eqref{eq:TaylorOp}. One gets, for example,  
$\gamma_1^{(1,0)}=\int dx_1 dt_1 \gamma^{(1,0)}_{x_1-x_i,t_1-t_n} (x_1-x_i)=\int dx dt\, \gamma^{(1,0)}_{x,t} x$.

Symmetries strongly restrict the coefficients of the Taylor expansion and, as we argue below, also by the fact that our system approaches an infinite-temperature steady state. We will work out the corresponding constraints for the dynamics in the following section.

\begin{table*}[t]
  \begin{center}
    \caption{Classification of all $255$ range-4 operators $B_i$.  Half of those operators ($128$) are not protected by hydrodynamics, $B_i|_\text{sym}=0$ and thus decay exponentially, see text and Fig.~\ref{fig:all_vanishing}. The remaining $127$ 
	operators, listed below, can be sorted into 9 sets of operators, which share the same long-time tail. 
	 $n_\text{op}$ gives the number of operators in each set.
The column labeled $R_{\pi \hat x}$  shows whether the operator $B_i|_\text{sym}$ is even or odd under rotations by $\pi$ around the $\hat x$. This determines whether it contains an even or odd power of $m$ within hydrodynamics. Similarly, the next column labeled $I_L$ identifies which operators $B_i|_\text{sym}$  are even or odd under inversion on a link. ``$-$'' indicates that the operator is neither even or odd. Even (odd) operators contain an even (odd) number of gradients. Based on these symmetries, one can find the leading operator in the field theory, listed in the next column. However, whether these operators obtain a finite expectation value or not depends on the initial condition.
Therefore, for the three initial conditions considered in our paper, the last three pairs of columns give an example of the most relevant operator with a finite expectation value and indicate how the observable decays in the long-time limit. 	 
	 The entry $0$ denotes that the expectation value of the operator vanishes by symmetry for the given initial 
	  condition. Note that the initial conditions also affect how fast a given operator decays. For example, $m^2$ (class 2 and class 3) is predicted to decay with $t^{-1/2}$ for the first initial condition but with $t^{-3/2}$ for the other two initial conditions. }
    \label{tab:table1}
    \begin{tabular}{l|l|c||c|c|c||c|c||c|c||c|c|} 
  &   \textbf{operators} & $n_\text{op}$ & $R_{\pi \hat x}$ & $I_L$ &\textbf{field theory} &
 \multicolumn{2}{c|}{\textbf{$|\psi_0^{(1)}\rangle$ (staggered)}}
& \multicolumn{2}{c|}{\textbf{$|\psi_0^{(2)}\rangle$ (x-pol.)}} 
& \multicolumn{2}{c|}{\textbf{$|\psi_0^{(3)}\rangle$ (current)}}\\
 &  &   &  & &leading &
 leading & decay &  leading & decay &  leading & decay \\
       \hline
 1&     $Z_i$               & 4  & $-1$ & $-$& $m$        & $(\nabla m)^3$ &    $t^{-4}$ & $-$ & 0 &$(\nabla m)^3$&   $t^{-4}$ \\
  2&      $Z_i Z_{i+1}$,\  \  $Z_i Z_{i+3}$ & 4 &1 &1&    $m^2$ & $m^2$ & $t^{-1/2}$   & $m^2$ &   $t^{-3/2}$ &  $m^2$ &   $t^{-3/2}$\\
   3&      $Z_i Z_{i+2}$ & 2  &1&$-$&    $m^2$ & $m^2$ & $t^{-1/2}$   & $m^2$ &   $t^{-3/2}$ &  $m^2$ &   $t^{-3/2}$\\
  4&       $Z_i Z_{i+n_1}Z_{i+m}$  & 4& $-1$&$-$ & $m^3$ & $(\nabla m)^3$ &    $t^{-4}$ & $-$ & 0 &$(\nabla m)^3$&   $t^{-4}$ \\
  5&      $Z_i Z_{i+1}Z_{i+2}Z_{i+3}$  & 1 & 1&1 & $m^4$& $m^4$ & $t^{-1}$ & $m^4$&  $t^{-2}$  & $m^4$&  $t^{-2}$ \\
  6&      $X_i X_{i+1}$,\ \ $Y_i Y_{i+1}$, $X_i X_{i+3}$,\ \ $Y_i Y_{i+3}$  & 16& 1&1 & $(\nabla m)^2$ &$(\nabla m)^2$ &  $t^{-3/2}$ &$(\nabla m)^2$ & $t^{-5/2}$ &$(\nabla m)^2$ &  $t^{-5/2}$\\
   &    $X_i X_{i+1}X_{i+2} X_{i+3}$,\ \ $Y_i Y_{i+1}Y_{i+2} Y_{i+3}$   & && && &  & &  & & \\
   &   $A_i B_{i+1} B_{i+2} A_{i+3}$, $A\neq B \in \{X,Y,Z\}$   & && && &  & &  & & \\
   7&      $X_i X_{i+2}$,\ \ $Y_i Y_{i+2}$  & 40& 1&$-$ & $ m \nabla m$ &$(\nabla m)^2$ &  $t^{-3/2}$ &$(\nabla m)^2$ &  $t^{-5/2}$ &$(\nabla m)^2$ & $t^{-5/2}$\\
   &   $A_i B_{i+1} A_{i+2} B_{i+3}$, $A\neq B \in \{X,Y,Z\}$   & && && &  & &  & & \\
   &   $A_i A_{i+1} B_{i+2} B_{i+3}$, $A\neq B \in \{X,Y,Z\}$   & && && &  & &  & & \\
   &    $X_i Y_{i+k} Z_{i+n}$,\ \ $Y_i X_{i+k} Z_{i+n}$   & && && &  & &  & & \\
  8&     $X_i Y_{i+1}$, \  \ $Y_i X_{i+1}$, \  \ $X_i Y_{i+3}$, \  \ $Y_i X_{i+3}$ & 12&  $-1$&$-1$& $ \nabla m $& $(\nabla m)^3$ & $t^{-4}$ & $-$ & 0  & $(\nabla m)^3$ & $t^{-4}$ \\
   & $X_i Z_{i+1} Z_{i+2} Y_{i+3}$,\ \  $Y_i Z_{i+1} Z_{i+2} X_{i+3}$   & && && &  & &  & & \\
   &   $Z_i X_{i+1} Y_{i+2} Z_{i+3}$,\ \  $Z_i Y_{i+1} X_{i+2} Z_{i+3}$  & && && &  & &  & & \\
 9&     $X_i Y_{i+2}$, \ \ $Y_i X_{i+2}$ & 44&   $-1$&$-$& $ \nabla m $& $(\nabla m)^3$ & $t^{-4}$ & $-$ & 0  & $(\nabla m)^3$ & $t^{-4}$ \\
   &    $X_i X_{i+n} Z_{i+m}$,\ \ $Y_i Y_{i+n} Z_{i+m}$  & && && &  & &  & & \\
   &    $X_i X_{i+k} X_{i+n} Y_{i+m}$,\ \ $Y_i Y_{i+k} Y_{i+n} X_{i+m}$   & && && &  & &  & & \\
    &    $X_i Y_{i+1} Z_{i+2} Z_{i+3}$, \ \   $Y_i X_{i+1} Z_{i+2} Z_{i+3}$  & && && &  & &  & & \\
    &    $Z_i Z_{i+1} X_{i+2} Y_{i+3}$, \ \   $Z_i Z_{i+1} Y_{i+2} X_{i+3}$  & && && &  & &  & & \\
    &    $Z_i X_{i+1} Z_{i+2} Y_{i+3}$, \ \   $Z_i Y_{i+1} Z_{i+2} X_{i+3}$  & && && &  & &  & & \\
   &    $X_i Z_{i+1} Y_{i+2} Z_{i+3}$, \ \   $Y_i Z_{i+1} X_{i+2} Z_{i+3}$  & && && &  & &  & & \\
  \end{tabular}
  \end{center}
\end{table*}

\subsection{Predicting long-time tails by classifying operators}\label{sec:predictions}

We have classified all 255 possible Hermitian operators $B_j$ which can be written as a product of $X_i$, $Y_i$, $Z_i$ (and the identity matrix) on four neighboring sites $i$, $i+1$, $i+2$, $i+3$. Table~\ref{tab:table1} shows how we can map the microscopic operators to objects in the field theory. 
For definiteness, we chose a convention where $i=4 n$ is always a multiple of 4. These 255 operators form (together with the identity) a complete basis of Hermitian operators on four neighboring sites.
Thus, they can be used to reconstruct the 4-site reduced density matrix. Using that $\tr(B_i B_j)=\delta_{i,j} 2^4$ for our operators, one obtains
\begin{align}\label{eq:rhored}
\rho^\text{red}=\frac{1}{2^4} \sum_{j=0}^{255} B_j \langle B_j \rangle.
\end{align}
where $B_0$ is the identity matrix.

To construct the table~\ref{tab:table1} and to predict long-time tails, we use a combination of seven different arguments. The first four are independent of the initial state, and the last three take properties of the initial state into account.

\subsubsection{Constraints arising from symmetries of the time evolution operator}\label{sec:constraint1}
(\textbf{i}) The effective field theory is rotation-invariant under spin rotations around the z-axis. Thus, only operators that are spin-rotation invariant can be described by this theory. Therefore, we symmetrize each operator $A$ using the generator of rotations $M_z=\sum Z_i$,
\begin{align}\label{eq:symmetrized}
A|_\text{sym}=\int_0^{2 \pi} \frac{d\alpha}{2 \pi} e^{i \alpha M_z/2}\, A \, e^{-i \alpha M_z/2} 
\end{align}
Three examples of symmetrized operators are
\begin{align}
X_i X_{i+1}|_\text{sym}&=\frac{1}{2} \left( X_i X_{i+1}+Y_i Y_{i+1}\right) \nonumber  \\
X_i Y_{i+1}|_\text{sym}&=\frac{1}{2} \left( X_i Y_{i+1}-Y_i X_{i+1}\right) \nonumber \\
X_i Y_{j} Z_{k}|_\text{sym}&=\frac{1}{2} (X_i Y_{j}-Y_i X_{j})Z_k.
\end{align}
 If $A|_\text{sym}=0$, the operator is not protected by conservation laws and is expected to decay exponentially. This happens if the total number of $X$ and $Y$ operators is odd and applies to $128$ of the $255$ operators considered by us. It is a crucial assumption of hydrodynamics that all operators, unprotected by slow modes, decay exponentially; this assumption will be tested numerically below. 
More generally, we expect an exponential decay for all expectation values of the type
\begin{align}\label{eq:expV}
\langle A-A|_\text{sym} \rangle \sim e^{-\Gamma t}.
\end{align}
Examples of exponentially decaying operators are $X_i$, $Y_i$, $X_i Y_{i+1} X_{i+2}$,  or $X_i Y_j + Y_i X_j$. 

(\textbf{ii}) As a second step, we consider how the operators transform under spin rotation by $\pi$ around the $x$ axis
\begin{align}
A &\to R_{\pi \hat x}^{-1} A R_{\pi \hat x}, \qquad  R_{\alpha \hat n}=e^{-i \alpha \sum \hat n \cdot \vec \sigma_i/2}\ .
\end{align}
If $A$ is even or odd under this transformation, its expansion into hydrodynamic modes will contain an even or odd power of $m$, respectively, as $m$ is odd under this symmetry. More precisely, also $\eta$ is odd under this symmetry, but we omit terms containing powers of $\eta$ in our analysis as $\eta$ correlators are local in time.

(\textbf{iii}) Next, we have to investigate how an operator $A|_\text{sym}$ transforms under the spatial link-inversion symmetry $I_L$ when we choose the link beween $i_0$ and $i_0+1$ as a center of inversion
 \begin{align}
I_L: \quad i \to (2 i_0+1) - i.
\end{align}
Note that spatial inversion centered on a site,  $i \to 2 i_0 -i$, is {\em not} a symmetry of our time evolution operator, as we apply the $XX$ and $YY$ operators in  Eq.~\eqref{eq:U} first on odd and only later on even links. 
If the operator $A|_\text{sym}$ is even or odd under this operation, it will contain an even or odd number of gradients within the effective field theory, respectively. Often, $A|_\text{sym}$ will, however, be neither symmetric nor antisymmetric under link inversion. For example, $X_1 X_3$ transforms as $X_1 X_3 \to X_2 X_4$ under link-inversion, but translation by a single site is not a symmetry of our time-evolution operator. In this case, we have to allow for both even and odd powers of gradients in the Taylor expansion.

(\textbf{iv}) Beyond symmetry, we have to take into account some special properties of our infinite-temperature fixpoint. Here, it is useful to consider a state with finite magnetization in the limit $t\to \infty$. As this is an infinite temperature state, all sites are uncorrelated and $\langle X_i\rangle = \langle Y_i\rangle=0$ while $\langle Z_i \rangle = \langle m\rangle$. We can use that to compute exactly expansion coefficients $\gamma_0^{(n)}$, the prefactors of $m^n$, in the Taylor expansion of operators in Eq.~\eqref{eq:taylor}.  Matching the left and the right side in \begin{align}
\lim_{t \to \infty} \langle A \rangle =\gamma_0^{(n)} \langle m^n\rangle_{T \to \infty} = \gamma_0^{(n)} \langle m\rangle^n
\end{align}
 we find 
 \begin{align}\label{eq:gammaZero}
\begin{array}{lll}
\gamma_0^{(n)}&=0 \quad & \text{for operators $A$  with at least one $X_i$ or $Y_i$ }\\
\gamma_0^{(n)}&=\delta_{n,m}\quad 
& \text{for operators $A=Z_{i_1}Z_{i_2} \cdot\dots \cdot Z_{i_m} $}
\end{array}
\end{align}
For example, the operator $X_i X_{i+1}$ can therefor {\em not} be proportional to $m^2$ but its Taylor expansion starts with $(\nabla m)^2$
\begin{align}
\langle X_i X_{i+1} \rangle \sim \langle (\nabla m)^2 \rangle.
\end{align}
 The same argument also shows that $Z_i Z_{i+n}\approx m^2$ with prefactor $1$ is independent of $n$, as we will check numerically below. 

\subsubsection{Constraints arising from symmetries of the initial state}\label{sec:constraint2}

The initial state and the time evolution operator will typically not have the same symmetries. In all cases considered by us, there is, however, a subgroup of symmetries shared by both the initial state and the time evolution operator. Such symmetries remain intact throughout time evolution and thus put extra strong constraints on the long-time dynamics.

(\textbf{v}) In our study, we consider only states that have a discrete translational invariance. A joint symmetry of initial state and time evolution is translations by either $2$ or $4$ sites, $i \to i+ 2 n_0$ or, equivalently, $x \to x+2 n_0 a$ in the field theory with $n_0=1,2$. Therefore, only expectation values which carry integer multiples of the momentum $\frac{ \pi}{n_0 a}$ can be finite.
Importantly, all operators with a finite momentum $k_0$ decay exponentially within our effective field theory, $\sim e^{-D k_0^2 t}$. Thus, to identify slowly decaying objects, we have to project each operator in the effective field theory to the zero-momentum sector. This is most easily done in momentum space. For example, $(\nabla m)^2$ projected to zero momentum gives
$-\sum_{k_1 k_2}  k_1 m_{k_1} k_2 m_{k_2} \delta(k_1+k_2)=\sum k^2 m_k m_{-k}$ where the projection is performed with the help of the $\delta$ function. In contrast, 
all operators which are total derivatives like $\nabla m^n$ or $\nabla (m^{n_1} \nabla^{n_2} m^{n_3})$ vanish after projection \cite{mukerjee2006statistical} and thus cannot have a finite expectation value. Consider, for example, $\nabla m^2$. In this case, projection gives $\sum_{k_1 k_2}  i (k_1+k_2) m_{k_1} m_{k_2} \delta(k_1+k_2)=0$.
We can also prove this more generally. Consider a combination of fields $A(x)$ which can be written as a total derivative, $A(x)=\nabla B(x)$. Any state with $\langle A\rangle=\nabla \langle B(x) \rangle=const.$ thus features an expectation value $\langle B(x) \rangle$ growing linearly in $x$, which is not possible in a translationally invariant system.
Using the rules for powercounting, Eq.~\eqref{eq:powercounting}, the projection onto the zero-momentum sector, and the symmetry arguments (ii) and (iii), we obtain the following table for leading operators in the zero-momentum sector
\begin{align}
 \begin{tabular}{c|c|c}
$R_{\pi \hat x}$& inversion $I_L$ & selected leading operators \\ \hline &&\\[-2mm]
even & even & $m^2 \text{ and } (\nabla m)^2$ \\
& odd & $ m^2(\nabla m) (\nabla^2 m)$ \\[2mm]
odd & even & $m$, $ m^3 \text{ and } m (\nabla m)^2$ \\
& odd & $(\nabla m)^3$
\end{tabular}
\end{align}

(\textbf{vi}) Our initial states have extra discrete symmetries jointly with the time evolution operator. For our arguments, we need the following symmetries
\begin{align}\label{eq.symminitial}
|\Psi_0^{(1)}\rangle:\qquad  & R_{\pi \hat x} I_L \ \ \text{and } R_{\alpha \hat z}, 0\le \alpha<2 \pi\nonumber \\
|\Psi_0^{(2)}\rangle:\qquad & R_{\pi \hat x} \ \ \text{and } I_L\nonumber \\
|\Psi_0^{(3)}\rangle:\qquad & R_{\pi \hat x} R_{(\pi/2) \hat z} I_L 
\end{align}
where, as above, $I_L$ is the inversion on a link and $ R_{\alpha\hat n}$ is the rotation of all spins around the axis $\hat n$. 
Any operator, either in the bare theory or the continuum, which is odd under this symmetry has to vanish.
For example, $\langle Z_1 Z_2 Z_3\rangle$ vanishes exactly for all times for the inital state $|\Psi_0^{(2)}\rangle$, as it is odd under rotations around the $\hat x$ axis by $\pi$.

 In contrast, $\langle Z_1 Z_2 Z_3\rangle$ is finite for the initial state $|\Psi_0^{(1)}\rangle$,
where initially  $\langle \Psi_0^{(1)}| Z_1 Z_2 Z_3|\Psi_0^{(1)}\rangle=1$.
Does this mean that $ \langle \int m^3(x) dx\rangle$ is finite in the continuum theory? The answer is negative, as evident from a symmetry analysis. $R_{\pi \hat x} I_L$ is a symmetry of the initial state $|\Psi_0^{(1)}\rangle$ and of the time evolution. Thus, the expectation value of any operator which is even under inversion but odd under $m\to -m$ has to vanish. This shows that $\left\langle \int m^3(x) dx \right\rangle=0$ or  $\left\langle \int m(x) (\nabla m(x))^2 dx \right\rangle=0$. Only terms that are odd in $m$ and have an odd number of gradients can have a finite expectation value. The same argument applies also for the initial condition $|\Psi_0^{(3)}\rangle$, where, however, initially $\langle \Psi_0^{(3)}| Z_1 Z_2 Z_3|\Psi_0^{(3)}\rangle=0$. For the three different initial conditions, we therefore conclude
\begin{align}
\langle  Z_1 Z_2 Z_3 \rangle^{(1)}&\sim (\nabla m)^3  \nonumber \\
\langle  Z_1 Z_2 Z_3 \rangle^{(2)}&=0 \nonumber \\
\langle  Z_1 Z_2 Z_3 \rangle^{(3)}&\sim (\nabla m)^3 .
\end{align}

(\textbf{vii}) Finally, there is one more property of the initial-state wave function, which we have to take into account when discussing long-time tails: the precise form of the correlations of the initial state.  As discussed in Eq.~\eqref{eq:correlationDecay1}, the long-time tails arise when initial-state correlations of hydrodynamic modes differ from their final state correlations.

Most importantly, the leading-order long-time tail $\sim 1/\sqrt{t}$ arises, as it takes a long time to build up the correlation $\langle m(x) m(x') \rangle \approx \gamma \delta(x-x')$ in the long-time limit \cite{Lux2014}. According to Eq.~\eqref{eq:correlationDecay1}, this leading contribution {\em vanish}, if the correlations in the initial state $C_{k=0}$ match correlations in the finial state $C^\eta_{k=0}/(2 D_0)$, as there is no need to redistribute magnetization in this case \cite{Lux2014}. 

The final state of our system is indistinguishable from an infinite temperature state in the thermodynamics limit, and therefore 
\begin{align}
\lim_{t \to \infty} \langle Z_i Z_{j} \rangle = \langle Z_i Z_{j} \rangle_{T=\infty}=\delta_{i,j}.
\end{align}
For our three initial states, we obtain 
\begin{align}
\langle \Psi_0^{(1)}|  Z_i Z_{j}| \Psi_0^{(1)} \rangle& = (-1)^{i+j} \nonumber  \\
\langle \Psi_0^{(2)}|  Z_i Z_{j}| \Psi_0^{(2)} \rangle& =\langle \Psi_0^{(3)}|  Z_i Z_{j}| \Psi_0^{(3)} \rangle =\delta_{i,j}.
\end{align}
Therefore, magnetization fluctuations of the initial state and final state are {\em identical} for the second and third initial conditions, and we conclude that leading-order long-time tails vanish in this case.

Subleading corrections arise, however, because initial and final states have different expectation values of, e.g., $X_i X_{i+1}$ and thus of  $(\nabla m)^2$. Therefore, the last term in Eq.~\eqref{eq:correlationDecay1} remains finite in this case,  $C''_{k=0}\neq \frac{C''^\eta_{k=0}}{2 D_0}$. Thus, we predict that 
\begin{align}
\langle Z_i Z_j \rangle^{(2,3)} \sim \frac{1}{t^{3/2}}
\end{align}
for the initial conditions $|\Psi_0^{(2)} \rangle$and $|\Psi_0^{(3)} \rangle$.

The same argument also applies to computations of operators with an overlap with $(\nabla m)^2$.  Using Eq.~\eqref{eq:correlationDecay1}, one can see directly that $\langle (\nabla m)^2\rangle \sim \frac{1}{t^{3/2}}$ if $C_{k=0} \neq C^\eta_{k=0}/(2 D_0)$, while $\langle (\nabla m)^2\rangle \sim \frac{1}{t^{5/2}}$ for  $C_{k=0} = C^\eta_{k=0}/(2 D_0)$.

Similarly, one can analyze the $m^4$ term. All expectation values $ \langle Z_i Z_j Z_k Z_l \rangle$ are identical in the initial and final state when starting from either $|\psi_0^{(2)}\rangle$ or $|\psi_0^{(3)}\rangle$ and therefore the leading long-time term $\sim \frac{1}{t}$ vanishes in this case. The initial and final state differ, however, by operators of the type $(\nabla m)^2 m^2$. The extra two gradients imply a decay with $1/t^2$.

As a concrete example of how the seven rules are applied, consider the operator
$X_i Y_{i+1}$.  Applying rules (i-iii), we find after symmetrization that the operator is odd in $m$ and odd under link-inversion
\begin{align}
 X_i Y_{i+1}|_\text{sym}=&\frac{1}{2} ( X_i Y_{i+1}- Y_i X_{i+1}) \nonumber\\
\sim &\nabla m  + (\nabla m) m^2 + \nabla^3 m + \nabla \partial_t m  \nonumber  \\
&+ (\nabla m)^3+ (\nabla m) (\partial_t m) m+ \dots \label{eq:currentOp}
\end{align}
The terms are ordered according to their scaling dimension using $m\sim t^{-1/4}, \nabla \sim t^{-1/2}, \partial_t \sim t^{-1}$, see Eq.~\eqref{eq:powercounting}. Thus $\nabla m$ appears to be the most relevant term.
This is, however, {\em not} correct, as we have not taken into account the projection to the zero-momentum section (rule (v)).
As all four terms in the second line of Eq.~\eqref{eq:currentOp} are total derivatives, they vanish when evaluated on a uniform state with momentum zero. Thus, scaling predicts (omitting again prefactors) that 
\begin{align}\label{eq:current_long}
\langle  X_i Y_{i+1} \rangle^{(1,3)} \sim \left\langle (\nabla m)^3\right\rangle+\left \langle  (\nabla m) (\partial_t m) m\right\rangle
\end{align} 
provided that this term is allowed by the symmetry of the initial state, which is the case for the initial conditions $|\psi_0^{(1)}\rangle$ and $|\psi_0^{(3)}\rangle$.

 \subsubsection{Overview over the resulting long-time tails}\label{sec:overviewRules}

 Applying the rules described above, we obtain a prediction for the leading power laws expected for an infinite system, which is shown in Table \ref{tab:table1}. Below, we comment on some entries in this table that are less intuitive, recapitulating some of our previous results.

 One of the, perhaps, most counter-intuitive results of our analysis is that the staggered magnetization $M_s$ and the uniform spin current $J$ have the same symmetries and correspond to the same operator in the effective field theory.
 \begin{align}
M_s&=\frac{1}{N} \sum_i (-1)^i \langle Z_i\rangle \sim J=\frac{1}{N} \sum_i \langle X_i Y_{i+1}-Y_i X_{i+1}\rangle \nonumber \\
&\sim \langle (\nabla m)^3 \rangle + \langle m (\nabla m) (\partial_t m ) \rangle.\label{eq:magCurr}
\end{align}
Both are odd under link-inversion $I_L$ and under  $R_{\pi \hat x}$, which maps $m$ to $-m$. As they have overlaps with the same operator in the effective field theory, a generic state with a staggered magnetization $M_s$ carries a finite spin current  $J$, and a generic state with a uniform current obtains a staggered magnetization, as we check numerically below. 
Note that in a conventional Heisenberg model, $M_s$ and $J$ differ by their transformation properties under site inversion. This is, however, not a symmetry of our system as we treat even and odd bonds in a different way during time evolution, Eq.~\eqref{eq:U}. 
This was necessary as we wanted to implement our model using only 2-bit gates. 
 
 A closely related observation is that the the two initial states, the staggered antiferromagnet  $|\Psi_0^{(1)}\rangle$ and the current carrying $|\Psi_0^{(3)}\rangle$ have very similar symmetries, $R_{\pi \hat x} I_L$ and  $R_{\pi \hat x} R_{(\pi/2)\hat z} I_L$, respectively, see  Eq.~\eqref{eq.symminitial}. The symmetries differ only by a rotation around the $\hat z$ axis. As our hydrodynamic theory only contains fields that are rotationally invariant with respect to this rotation, the two initial states are identical from the point of view of hydrodynamics.  Thus, we obtain the same leading operators in Table \ref{tab:table1} (although not the same power laws, see below).
 
Next, we discuss how the leading long-time tail is computed for operators like $J$ or $M_s$, which have a finite overlap with $(\nabla m)^3$. In the effective field theory, the presence of a current or a staggered magnetization is encoded by 
a $t=0$ correlation function, see rule (vi)
\begin{align}
\langle m_{k_1} m_{k_2} m_{k_3}  \rangle\approx &\\
 & \hspace{-1cm}\delta(k_1+k_2+k_3)i \left(c_1  k_1 k_2 k_3+c_2  (k_1^3+k_2^3+k_3^3)\right) \nonumber
\end{align}
where only the leading term in a Taylor expansion is shown and we used that the initial states have a symmetry ($R_{\pi \hat x} I_L$ or  $R_{\pi \hat x} R_{(\pi/2)\hat z} I_L$) which enforces that
terms odd in $m$ must also be odd in $k$. Here, terms linear in $k$ cannot show up as they correspond to total derivatives, see rule (v).
Repeating the calculation of Eq.~\eqref{eq:correlationDecay1} using that noise terms do not produce odd-in-m terms, one obtains
\begin{widetext}\label{t4decay}
		\begin{align}
		\langle (\partial_x m(x,t))^3\rangle & 
		\sim \int dk_1 dk_2 d k_3  i k_1 k_2 k_3 G_{k_1}(t)G_{k_2}(t)G_{k_3}(t) \delta(k_1+k_2+k_3)i c_1  k_1 k_2 k_3 \sim \frac{c_1}{t^4}.
		\end{align}
\end{widetext}

Finally, we want to finish our discussion of table \ref{tab:table1} by pointing out that the same type of operator in the effective field theory can lead to different power laws. While the operators $m^2$, $(\nabla m)^2$, and $m^4$ decay with exponents $t^{-1/2}$, $t^{-3/2}$, 
and $t^{-1}$ for the first initial condition, their decay is faster by a factor $1/t$ for the two other initial conditions.
The origin of this behavior is explained under rule (vii): the initial and final state are characterized by the same leading-order fluctuations of the conserved quantities for both $|\Psi_0^{(2)}\rangle$ and $|\Psi_0^{(3)}\rangle$ and therefore the leading long-time tail cancels.

\section{Numerical results and comparison to hydrodynamics}\label{sec:num}

\subsection{Computation of long-time tails}

In our numerics, we consider systems with up to $N=28$ spins, where one can compute the time evolution defined in Eq.~\eqref{eq:timeevolution}   `brute force'. After each time step, we compute the expectation value of all 255 observables, which can be defined on four neighboring sites.
We follow the time evolution for typically $300$ time steps starting from the initial conditions defined in Eq.~\eqref{eq:initial}. 
For longer times, finite-size effects dominate, and a finite-size background noise dominates most expectation values.

For the following comparison of numerics and analytics, we would like to emphasize some limitations of our study. First, hydrodynamics is only valid in the long-time limit. More precisely, we are interested in the behavior that occurs in the thermodynamic limit, implying that we want to study times short compared to the time scale of diffusion through the full system, $t\ll N^2 /D_0$. In contrast,  Ref.~\cite{Maceira2024} focuses on the opposite limit, where an exponential decay of observables is expected.

 The extremely slow decay, $1/\sqrt{t}$, of long-time correlations makes fits difficult as
subleading terms are only suppressed by relative factors of $1/\sqrt{t}$. Thus, any observable has large subleading corrections.
In practice, large corrections imply that we are unable to determine exponents by fitting. Our goal is therefore not to `prove' numerically that 
predictions from hydrodynamics are valid. Instead, we have a more modest goal: we show that the numerical data is consistent with the expectation from hydrodynamics. We can show this consistency for a huge set of observables. In a few cases, however, non-universal prefactors of the long-time tails are accidentally so small that possible power-law tails become indistinguishable from a pure exponential decay. After this disclaimer, we can now start to explore the validity of hydrodynamics in our model.

	\begin{figure}[t]
		\includegraphics[width=1. \columnwidth]{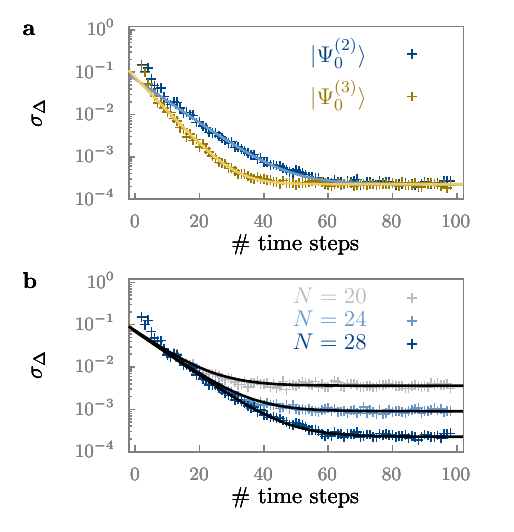}
		\caption{\label{fig:decayNorm} 
Only operators symmetric under rotations around the $z$ axis are protected by hydrodynamics and decay slowly, while the difference $\langle A-A|_\text{sym}\rangle$ always decays exponentially. To test the assumption underlying our hydrodynamic analysis,  we plot the standard deviation $\sigma_\Delta$ of this quantity averaged over all observables as defined in Eq.~\eqref{eq:operatorNormDiff}. The solid lines are fits to $a e^{- t/\tau}+b \, 2^{-N/2}$, Eq.~\eqref{eq:fit}, with   $b \approx 3.7$. The offset arises from the finite-size effect and is approximately the same for the two initial conditions. Panel \textbf{a}: For the  initial conditions $|\psi_0^{(2)}\rangle$ and $|\psi_0^{(3)}\rangle$, we obtain from the fit $\tau\approx 8.3$ and $\tau\approx 5.4$, respectively (for the initial condition $|\psi_0^{(1)}\rangle$ one obtains $\sigma_\Delta=0$ by symmetry).
The exponential decay also shows that the reduced density matrix of the subsystem approaches its symmetrized version with exponential precision, see Eq.~\eqref{eq:operatorNormDiff}. Panel \textbf{b}: The fit obtained from panel {\bf a}  describes (without readjusting fitting parameters) very well the data for $N=20$ and $N=24$, thus confirming that the offset is exponentially suppressed for $N\to \infty$. Parameters for all figures: $\alpha=2$, $\beta=0.25$, $\gamma=1$, and $N=28$ with the exception of  two curves in panel {\bf b}.}
	\end{figure}
\subsection{Exponentially fast decaying correlations}
The most important assumption underlying the mapping to hydrodynamics is that all degrees of freedom, which are {\em not} protected by hydrodynamics, decay exponentially on a finite time scale which is independent of system size. This applies to all operators $A - A|_\text{sym}$, see Eq.~\eqref{eq:expV} and Eq.~\eqref{eq:symmetrized}.

To check this numerically, we compute this quantity for all 255 operators $B_i$ considered by us and calculate the standard deviation
\begin{align}\label{eq:operatorNormDiff}
\sigma_{\Delta}&=
\left(\frac{1}{2^4}\sum_{i=1}^{255} \left\langle B_i - B_i|_\text{sym}\, \right\rangle^2\right)^{1/2} \nonumber \\
&=\bigl|\!\bigl|\  \rho^\text{red}-\rho^\text{red}|_\text{sym}\ \bigl|\!\bigl|.
\end{align} 
If and only if $\langle B_i - B_i|_\text{sym}\rangle$ decays at least exponentially for all $i$, we can expect an exponential decay of $\sigma_\Delta$.
Using Eq.~\eqref{eq:rhored}, this quantity can also be identified with the operator norm (defined by $|\!| A |\!|=(\frac{1}{2^4} \sum |A_{ij}^2|)^{1/2}$) of the difference of the reduced density matrix $\rho^\text{red}$  of the subsystem and its symmetrized version.

\begin{figure}[t]
		\includegraphics[width=1.0 \columnwidth]{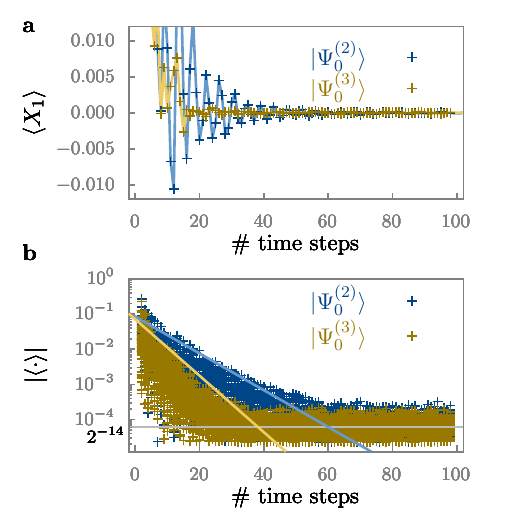}
		\caption{\label{fig:all_vanishing} Exponential decay of operators not protected by hydrodynamic modes.
                   Panel \textbf{a}: Time dependence of $\langle X \rangle$ for two different initial conditions, $|\Psi_0^{(2)}\rangle$ and $|\Psi_0^{(3)}\rangle$.
Panel \textbf{b}: Decay of 128 operators which are odd under
		rotations by $\pi$ around the $z$ axis and thus expected to decay exponentially. We plot the absolute value of the expectations for each operator 
		in a log plot for two different initial states. Note that for the third initial state considered in this
		paper (staggered magnetization), the expectation values are exactly zero at all times.  At long times, the expectation values fluctuate around zero with amplitudes of order $2^{-N/2}$ (gray line).
		The operators are decaying more slowly for the fully magnetized initial state (blue crosses), as indicated by the light blue line, compared to the state (yellow crosses), as indicated by the light yellow line. The solid blue and yellow line denotes an  exponential decay with the decay rates obtained from the fit in Fig.~\ref{fig:decayNorm}{\bf a}.}
	\end{figure}
	In Fig.~\ref{fig:decayNorm}, we show this quantity. It decays exponentially but obtains a finite value for $t\to \infty$. To obtain an estimate for this offset, we consider the Lehmann representation of an operator $A$
\begin{align}\label{eq:lehmann}
\langle A(t)\rangle=\sum_{n m} \langle \Psi_0 | n \rangle \langle n |A| m \rangle \langle m |  \Psi_0 \rangle e^{i (\epsilon_n -\epsilon_m)t}.
\end{align}
The long-time limit is obtained from the fluctuations of the matrix elements \cite{srednicki1999approach}. For a completely random system with a Hilbert space of size $2^N$, each of the three matrix elements is estimated to be of order $1/\sqrt{2^N}$. The sum contains $(2^N)^2$ terms of random sign. Therefore, we can estimate the size of the sum to be of order
$\sqrt{(2^N)^2} /(\sqrt{2^N})^3 \sim 2^{-N/2}$ with a random sign,
\begin{align}\label{eq:background}
\langle A(t)\rangle\sim \pm 2^{-N/2} \quad \text{for } t\to \infty.
\end{align}
Therefore, we expect for large $t$
\begin{align}\label{eq:fit}
\sigma_\Delta \approx a \, e^{-t/\tau} + b\, 2^{-N/2}.
\end{align}
The fit of  Fig.~\ref{fig:decayNorm}, which was obtained for system size $N=28$ and could be applied to $N=20$ and $N=24$ without readjusting the fitting parameters, confirms this result with $b\approx 3.7$ independent of $N$. The value of $\tau$ depends on the initial conditions.
For $|\Psi_0^{(2)}\rangle$ and $|\Psi_0^{(3)}\rangle$, we obtain $\tau\approx 8.3$ and $\tau \approx 5.4$, respectively. Apparently, the slowest exponentially decaying mode in the system with $\tau\approx 8.3$ is only activated for the state $|\Psi_0^{(2)}\rangle$, see below.

From the 255 operators investigated by us, 128 have the property that they are odd under rotation around the $z$ axis by $\pi$, $ e^{-i \pi M_z/2} B_i \, e^{i \pi M_z/2}=-B_i$ and therefore $B_i|_\text{sym}=0$.  These operators only take a finite expectation value if the initial condition is not rotationally invariant, i.e., for  $ |\psi_0^{(2)}\rangle$ and $ |\psi_0^{(3)}\rangle$.  In these cases, we expect
\begin{align}
\langle B_i(t) \rangle \approx \sum_j a_j e^{-\Gamma_j t} \cos(\omega_j t +\phi_j)+O(2^{-N/2}) 
 \end{align} 
where the exponentially small offset is not a constant but a randomly oscillating function of time.

From our observables, we find that $\langle X_i \rangle$ decays slowest for the initial condition $|\Psi_0^{(2)}\rangle$, which has a finite net magnetization in the $x$ direction. Apparently, its decay is the slowest exponential decay mode in the system and is thus responsible for the longest relaxation time with $\tau\approx 8.3$. 
 Fig.~\ref{fig:all_vanishing}a shows that this slow decay occurs in combination with an oscillation.
 In Fig.~\ref{fig:all_vanishing}b, we analyze the decay of the expectation values of all 128 observables for which we predict an exponential decay in a logarithmic plot. The solid lines in Fig.~\ref{fig:all_vanishing}b describe an exponential decay with the prefactor obtained from the fit to $\sigma_\Delta$, Eq.~\eqref{eq:fit} to confirm that all quantities decay exponentially and that the fitted relaxation times indeed describe the slowest of the exponentially decaying modes. Thus, our best estimate for the local equilibration time is $\tau_\text{loc}=\max \tau_i \approx 8.3$, but strictly speaking, $\sigma_\Delta$ and the quantities analyzed in this section only probe operators perpendicular to the space of $M_z$-conserving, symmetrized operators.

In conclusion, our numerical results fully confirm that operators not protected by hydrodynamics decay exponentially.

\subsection{Long-time tails in $\langle Z_i Z_{i+n}\rangle$}\label{sec:ZZsub}
\begin{figure}[t]
		\includegraphics[width=1.0 \columnwidth]{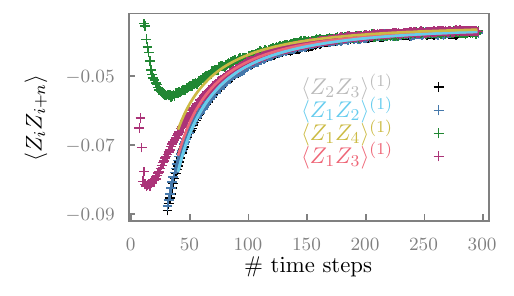}
		\caption{\label{fig:ZZfit} Evolution of $\langle Z_i Z_{i+n} \rangle$ after a quench as a function 
		of time for the initial state $|\psi_0^{(1)}\rangle$ (staggered magnetization). Dashed curves: Fitting function $f(t)=\frac{c_0}{N}+\frac{c_1}{\sqrt{t}}+\frac{c_2}{t}+\frac{\alpha_i}{t^{3/2}}$, Eq.~\eqref{eq:asym}, 
		with $\frac{c_0}{N}=-0.04$, $c_1=0.2$, $c_2=-2.63$, 
		$\alpha_{\langle Z_1 Z_{2} \rangle}=- 1.05$, $\alpha_{\langle Z_2 Z_{3} \rangle}=- 1.08$,
		$\alpha_{\langle Z_1 Z_{3} \rangle}=  0.11$, 
		and $\alpha_{\langle Z_1 Z_{4} \rangle}= 2.03$. Here, $c_1$ and $c_2$ are set equal for all 
		curves, see text. $c_1$ is the standard hydrodynamic long-time tail, $c_2$ arises from non-linear corrections in the diffusion equation. In contrast, the $\sim 1/t^{3/2}$ 
		dependence can originate from different modes (e.g., higher-order gradient terms) that have distinct prefactors. The fit indicates that the numerical result is consistent with the predictions from hydrodynamics; however, due to subleading terms and finite-size effects, it is not possible to reliably extract exponents from the numerical data.}
	\end{figure}

Next, we investigate the decay of the slowest mode in the system, $\langle Z_i Z_{i+n}\rangle$.
$\langle Z_i Z_{i+n}\rangle$ is expected to decay asymptotically proportional to $1/\sqrt{t}$, as it has a finite overlap with $m^2$ of the hydrodynamic theory, see Eq.~\eqref{eq:m2} . Fig.~\ref{fig:ZZfit} shows that these expectation values indeed decay extremely slowly for an initial state with staggered magnetization, $|\Psi_0^{(1)}\rangle$. Even after $300$ time steps, the observables have not yet reached a steady state. The expected decay exponent $1/2$, however, cannot be directly determined by a fit as subleading corrections are too large. 
Subleading corrections to the  $1/\sqrt{t}$ behavior come from three different sources: finite size effects, non-linearities in the hydrodynamics field theory, and higher-order terms in the gradient expansion, see Eq.~\eqref{eq:correlationDecay1} and Refs.\cite{Michailidis2024, delacretaz2024nonlinear}.

For a finite-size system with periodic boundary conditions and $t \gg N^2/D_0$, the diffusion equation predicts an exponential decay of all correlations $\sim e^{-D_0 (2 \pi/N)^2 t}$. We have not attempted to explore this long-time behavior in our numerics as our study aims to investigate long-time tails in the thermodynamic limit, $N \to \infty$. Nevertheless, for fits we have to take into account a possible finite offset. This offset is expected to be of order $1/N$, which follows from scaling arguments: time scales proportional to the square of length in a diffusive system. Thus, we expect a $1/N$ offset for an observable decaying with $1/\sqrt{t}$. The leading nonlinear term in the formula for the current, Eq.~\eqref{eq:current}, $D_2 m^2 \nabla m$, is suppressed by a factor $m^2$ relative to the leading term $-D_0 \nabla m$.
As $m^2$ scales as $1/\sqrt{t}$, this gives rise to a relative correction of order $1/\sqrt{t}$, leading to a correction of order $1/t$ to the expectation value of  $\langle Z_i Z_{i+n}\rangle$. Finally, higher-order corrections arise from higher-order gradient terms either in the operator expansion or in the correlations of the initial state, see Eq.~\eqref{eq:correlationDecay1}. 
Therefore, we conclude that
\begin{align}\label{eq:asym}
\langle Z_i Z_{i+n}\rangle^{(1)} \approx \frac{c_0}{N}+\frac{c_1}{\sqrt{t}}+\frac{c_2}{t}+\frac{\alpha_n}{t^{3/2}}+O\left(\frac{1}{t^2}\right).
\end{align}
Here, the prefactors $c_0$, $c_1$, $c_2$ are expected to be independent of the distance $n$ (for small $n$), as they arise from zeroth-order gradient expansion and as there is a single slow mode in the field theory, the fluctuations of the local magnetization, $m^2$, which is responsible for this long-time tail. Physically, this long-time tail arises because it takes a long time for these fluctuations to change through diffusive processes.
In contrast, $\alpha_n$ is different for different $n$. When performing a gradient expansion, the prefactor of the $(\nabla m)^2$ term in Eq.~\eqref{eq:taylor} depends on $n$, but one also has to take into account a contribution due to $m \partial_t m$. Both terms contribute to the $1/t^{3/2}$ dependence at long times.
In Fig.~\ref{fig:ZZfit}, the fits show that our numerical data is fully consistent with the prediction of Eq.~\eqref{eq:asym}.
We expect that the long-time tails {\em vanish} if the correlations in the initial state match correlations in the final state (rule vi). 
Magnetization fluctuations of the initial state and final state are {\em identical} for the second and third initial conditions, and we conclude that $c_0=c_1=c_2=0$ in Eq.~\eqref{eq:asym} in this case.
Indeed,  $\langle Z_i Z_{j} \rangle$ takes much smaller numerical values in this case (not shown) and decays to a good approximation with $1/t^{3/2}$. 

	\begin{figure}[t!]
		\includegraphics[width=1.0 \columnwidth]{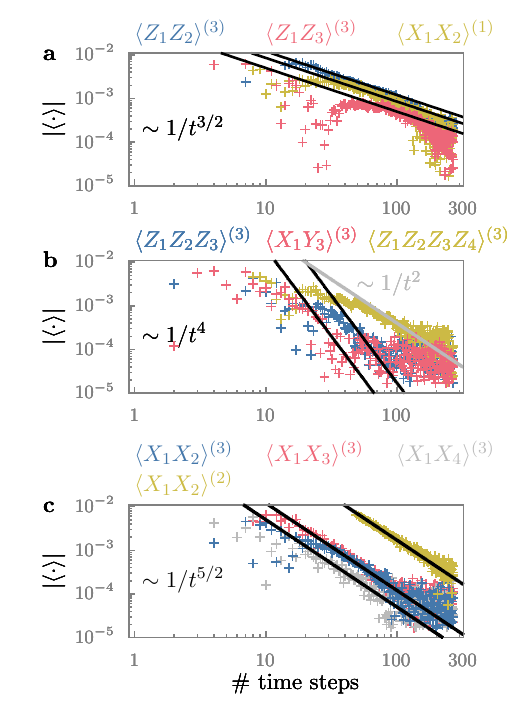}
		\caption{\label{fig:compare_all_log}
		Long-time decay of selected observables.
For the initial state
$|\psi_0^{(3)}\rangle$, we show one example for each of the nine symmetry classes shown in Tab.~\ref{tab:table1} except for classes 1 and 8, which are displayed in Fig.~\ref{fig:staggeredM_current}. Furthermore, we show operators of class 6 for all three initial conditions.
As shown in this log-log plot of the absolute value of the expectation values, all operators decay slowly as a power law. 
Panel \textbf{a}: Operators of class  6 (dark yellow) decay for $|\psi_0^{(1)}\rangle$  as $(\nabla m)^2\sim 1/t^{3/2}$, as indicated by the straight line (no fit). The same power law is also obtained for operators of classes 2 and 3 (blue and red) for the initial condition 
 $|\psi_0^{(3)}\rangle$. In this case, the observable is proportional to $m^2$ rather than $(\nabla m)^2$ , but the prefactor of the leading powerlaw $1/t^{1/2}$ cancels   as $C_{k=0} = C^\eta_{k=0}/(2 D_0)$ for this initial condition. Thus, one expects $1/t^{3/2}$ instead.
Panel \textbf{b}: Operators of class 5 (dark yellow) for $|\psi_0^{(3)}\rangle$ decay as  $ m^{4}\sim1/t^{2}$ indicated by the gray line (no fit). 
Operators of class 4 and 9 are predicted to decay as $(\nabla m)^{3}\sim 1/t^{4}$, indicated by the black lines (no fit), see Fig.~\ref{fig:staggeredM_current} for a discussion of other $1/t^4$ predictions.
Panel \textbf{c}:	This panel collects operators which decay as  $1/t^{5/2}$ (black lines). This exponent is predicted for
the initial conditions $|\psi_0^{(2)}\rangle$ and $|\psi_0^{(3)}\rangle$ for operators of class 6 and 7.  Note that the same operators decay with $1/t^{3/2}$ for the first initial condition as shown in panel {\bf a}.
Expectation values of all other operators and including for other initial states can be retrieved at Zenodo~\cite{Zenodo}.}
	\end{figure}

\subsection{Other slowly decaying observable}

	\begin{figure}[t]
		\includegraphics[width=1 \columnwidth]{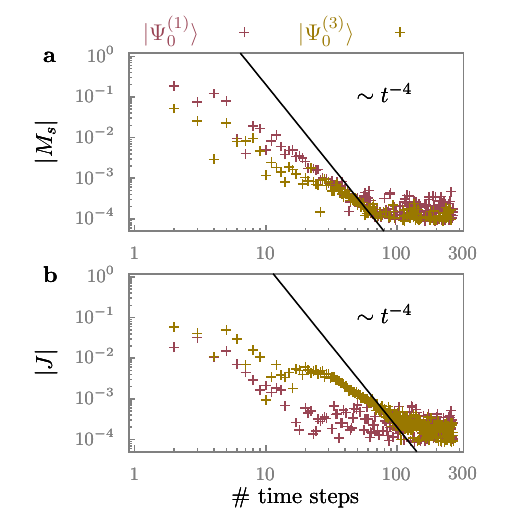}
		\caption{\label{fig:staggeredM_current} Decay of \textbf{a} staggered magnetization $M_s$ and \textbf{b} the current $J$ for the two initial conditions $|\Psi_0^{(1)}\rangle$ and $|\Psi_0^{(3)}\rangle$, where the two observables initially take the value $M_s=1$, $J=0$ and $M_s=0$, $J=1$, respectively. Our symmetry analysis and hydrodynamics predict that these quantities are linked to each other and decay with $1/t^{4}$, Eq.~\eqref{eq:magCurr}, shown as the solid lines. The numerical data appears to decay more slowly than predicted in this case (see text).
} 
\end{figure}

We have analyzed the decay of all observables in Table \ref{tab:table1} and show selected results from all classes 
in Fig.~\ref{fig:compare_all_log} and Fig.~\ref{fig:staggeredM_current}.  The complete set of data can be retrieved from Zenodo \cite{Zenodo}, see data availability statement. Below, we show that for most observables, we find a very good agreement with predictions from hydrodynamics, with the notable exception of some current-like operators, which appear to decay more slowly than predicted.

Fig.~\ref{fig:compare_all_log} shows examples from 7 of the 9 classes, where we find powerlaws of the form $1/t^{3/2}$, $1/t^{5/2}$, and $1/t^2$.  In these cases, we find that the numerical results are fully consistent with the predictions from hydrodynamics. This also includes cases like $\langle X_1 X_2\rangle$, where the power law depends on initial conditions. While the first initial condition (dark-yellow curve in panel {\bf a}) is expected to decay as $1/t^{3/2}$, the decay is much faster, $1/t^{5/2}$, for the two other initial conditions.

The situation is less clear for all cases where  Table \ref{tab:table1} predicts a $1/t^4$ decay of the operator, which includes operators like the staggered magnetization and the spin current. We recall, that $|\Psi_0^{(1)}\rangle$ has a staggered magnetization as an initial condition while $|\Psi_0^{(3)}\rangle$ carries a large spin current. A first prediction of our field-theoretical analysis is that both quantities couple to the same term in the effective field theory, $(\nabla m)^3$. Thus, after a short time, $|\Psi_0^{(1)}\rangle$ should carry  a current and $|\Psi_0^{(3)}\rangle$ a staggered magnetization, respectively. Fig.~\ref{fig:staggeredM_current}, which plots both quantities for both initial conditions, shows that this is, indeed, the case. Furthermore, we expect a decay of this quantity with $1/t^4$. Remarkably, the numerical data is not consistent with this expectation, but the data appears to decay more slowly. For example, the decay of the current for the initial condition 
$|\Psi_0^{(3)}\rangle$ is much better described by $1/t^2$ and the decay of the staggered magnetization for $|\Psi_0^{(1)}\rangle$ is consistent with a $1/t^{5/2}$ relation.

What can be the origin of this apparent discrepancy? First, it is not surprising that large exponents like $1/t^4$ are difficult to fit as they decay rapidly, the signal is thus very weak, and hence issues like finite-size effects become more important. We note that a recent study of a spin-1 model also found apparent numerical discrepancies for rapidly decaying cumulants, which were attributed to fine-size effects \cite{wang2025}.
An unexpected numerical result was also reported in Ref.~\cite{Maceira2024}, where a diffusive quantum system with energy conservation was studied. Here, the authors identified a relaxation rate in the finite system proportional to $1/N$ instead of $1/N^2$ expected from the diffusion equation. It is unclear whether these numerical observations are related to each other.

We have also tried to identify loopholes in our field theoretical argument, e.g., searching for extra contributions not described by Eq.~\eqref{t4decay}, which arise from non-linearities or finite-size effects, but we have 
not been successful in identifying a possible mechanism of a slower decay on the field theoretical side. We have to leave that as an open, unresolved question.

\section{Conclusions and Outlook}\label{sec:conclusion}
For large classes of self-equilibrating many-particle quantum systems, the long-time dynamics of local observables is expected to be described by fluctuating hydrodynamics, i.e., by a classical field theory describing hydrodynamic modes and their fluctuations. Importantly, corrections to this theory decay exponentially, where the equilibration time remains finite in the thermodynamic limit. In contrast,  hydrodynamic modes relax with power laws even in translationally invariant systems. We have put this theoretical framework to the test by (i) developing a set of analytical predictions from hydrodynamics for local observables and (ii) comparing these predictions with our numerical simulations for a large set of local observables. 

For most observables, we find that the classical effective field theory accurately describes the long-time dynamics of the quantum system. As discussed above, however, a set of observables related to spin current and staggered magnetization appears to decay more slowly than expected. We have not found a satisfactory explanation for this apparent discrepancy. It could be a numerical artifact of our finite-size study, indicate a technical mistake in our theoretical analysis, or suggest the existence of another slow mode that is not captured by the effective field theories. Extra slow modes may, for example, arise in the proximity of some integrable point \cite{Jung2006,Jung2007}. To resolve this question, it would be useful to extend our numerical study to larger system sizes and also to other models.

	\emph{Data availability:} We note that all raw data can be retrieved at Zenodo \cite{Zenodo}. The data includes the time series of the expectation values of all 255 operators. Each time series has 301 time steps. We provide the data for all three initial conditions and system sizes $N=20$, $N=24$, and $N=28$. A Mathematica notebook is also provided to create plots of the paper and beyond. This notebook allows for exploring the classification predicted in Tab.~\ref{tab:table1} and replaces an appendix with $3\times 3\times 255$ plots. 

	\emph{Acknowledgements:}	We thank Andreas L\"auchli and Ivo Maceira for useful discussions and, especially, Luca Delacretaz, who pointed out a major mistake in an earlier version of the manuscript. We acknowledge funding from the German Research	Foundation (DFG) through CRC 183 (project number 277101999, A01 and B02) and – under Germany’s Excellence Strategy – by the 	Cluster of Excellence Matter and Light for Quantum 	Computing (ML4Q) EXC2004/1 390534769. 
	\bibliography{literatureThermalization.bib}

	\end{document}